\documentclass[a4paper,11pt]{article}
\usepackage[british]{babel}
\usepackage{csquotes}
\usepackage{debug}
\usepackage{authblk}
\usepackage{hyperref}
\usepackage{amsmath}
\usepackage{amssymb}
\usepackage{amsfonts}
\usepackage{enumitem}
\numberwithin{equation}{section}

\newcommand{\eom}{{e.o.m}}
\newcommand{\wrt}{{w.r.t.}}
\newcommand{\oh}{ \frac{1}{2} }
\newcommand{\R}{ \mathbb{R} }

\newcommand{\vx}{{\vec x}}
\newcommand{\vz}{{\vec z}}
\newcommand{\xp}{{x^+}}
\newcommand{\xm}{{x^-}}

\newcommand{\hphi}{{\hat \phi}}
\newcommand{\thphi}{{\widetilde {\hat \phi}} }
\newcommand{\thsphi}{{\widetilde {\hat \phi^*}} }
\newcommand{\kv}{{k_v}}
\newcommand{\kvN}[1]{{k_{#1\,v}}}

\newcommand{\ein}{{\zeta}}
\newcommand{\eph}{{e_{ph}}}
\newcommand{\pv}{{p_v}}
\newcommand{\tpsi}{{\tilde \psi}}
\newcommand{\tPsi}{{\tilde \Psi}}
\newcommand{\cD}{{\cal D}}

\usepackage[
  backend=biber,
  style=numeric,
  citestyle=numeric-comp,
  bibstyle= numeric,
  sorting=none 
  ]{biblatex}
\title{Light-Cone Quantization a of Scalar Field on Time-Dependent Backgrounds}

\author{Andrea Arduino\footnote{andrea.arduino@to.infn.it}
}
\author{Igor Pesando\footnote{ipesando@to.infn.it}
}
\affil
{%
  Dipartimento di Fisica, Universit\`{a} di Torino \authorcr
  and I.N.F.N. -- sezione di Torino \authorcr
  Via P.\ Giuria 1, I-10125 Torino, Italy
}

\hypersetup
{%
  pdfauthor={Andrea Arduino, Igor Pesando},
  pdfsubject={String Theory},
  pdfkeywords={string theory, evolution, time dependent background},
}

\begin{document}

  \maketitle

  \begin{abstract}

We discuss what is light-cone quantization on a curved spacetime also
without a null Killing vector.
Then we consider as an example
the light-cone quantization of a scalar field on a
background with a Killing vector and the connection with the
second quantization of the particle in the same background.
It turns out that the proper way to define the light-cone quantization
is to require that the constant light-cone time hypersurface is null
or, equivalently, that the particle Hamiltonian is free of square roots.
Moreover, in order to quantize the scalar theory
it is necessary to use not the original
scalar rather a scalar field density, i.e.
the Schrödinger wave functional depends on a scalar
density and not on the original field.
Finally we recover this
result as the second quantization of a particle on the same background, where
it is necessary to add as input the fact that we are dealing with a scalar density.

  \end{abstract}
  
  \newpage

  \tableofcontents

\section{Introduction}
In this paper, inspired by the use in string theory of the light-cone
gauge in time-dependent backgrounds (see for example
\cite{Amati:1988ww}), we examine the light-cone
quantization of a scalar field in a light-cone time-dependent
background and
how it is related to the second quantization of the particle in the same
background.
For some reviews on light-cone quantization see
\cite{Heinzl:2000ht, Martinovic:2007nee}.

The first point which we discuss is what is meant by light-cone
evolution and then  quantization on a curved
time-dependent background, even if it does not admit a
null Killing vector.
In flat space there are different possible definitions which are equivalent but on a curved background they result in different outcomes.
The definition we choose is the one used implicitly in string theory
literature, i.e. the gauge in which the particle Hamiltonian does not
contain a square root.
The reason is that it gives the simplest form of the Hamiltonian but with
a drawback: the evolution may be in ``space-like'' directions in
certain points.
However this definition is equivalent to requiring that the equal time
hypersurfaces are null surfaces.


Then we consider the Hamiltonian formulation and the quantization of a
scalar in the light-cone of a generic time-dependent background which
admits a light-cone evolution according to the previous definition.
This computation makes clear that when we are studying light-cone quantization
in time-dependent backgrounds the field we quantize is not a scalar field
 but a scalar density.

Finally we can recover the previous results from the second
quantization of the particle only when we know that we are dealing with a scalar density.
Furthermore the comparison between the field theory and the particle formalism makes clear that the modulus squared of the ``wave function" of the light-cone formalism has to be interpreted as charge density and not as probability density.  

\section{Light-Cone Evolution in a Time-Dependent Background
}

In flat space and with the usual coordinates
it is possible to define the light-cone evolution in
one of the following ways:
\begin{enumerate}[label=\arabic*)]
\item
  Propagation in a null direction, i.e the Hamiltonian is the
  generator associated with a null Killing vector;
\item
  The constant time hypersurface is null, i.e. the tangent space
  contains a null vector; 
\item
  The gauge choice of worldline diffeomorphisms which yields the gauge
  fixed Hamiltonian  without square roots;
\item
  The choice of Hamiltonian which maximize the number of kinematical
  generators of the Poincar\'e algebra.
\end{enumerate}

While all these points of view give the same result in flat
space, the same is not true in a curved time-dependent background.
First of all let's notice that in such a background there is no global Poincar\'e algebra\footnote{When this happens it is somewhat intuitive that maximizing the
number of kinematical generators is equivalent to choosing
$p_u = \frac{\vec p^{\,2} +m^2}{2 p_v}$ as the particle Hamiltonian without square roots.}, which means we can't rely on 4).
We are therefore left to examine the consequences of the other three
possible definitions.

The first one is technically flawed because what really matters in the
evolution is not the direction of propagation but the ``orthogonal''
direction to the constant time hypersurface and the lapse, which measures the proper time between two hypersurfaces.
We write ``orthogonal'' since only when the hypersurface is spacelike
the propagation direction is orthogonal; on the contrary this is not true in the lightlike case, even if
there is still a well defined way of computing it as discussed below after equation \eqref{eq:normal_in_lc}.
Moreover the constant time hypersurface may be  neither completely
spacelike
nor completely lightlike. Let's therefore analyse definition 2) through some basic examples.

Consider the 2D Minkowski spacetime and change coordinates as
$t=\eta, x=\chi+ \beta \eta$, so that the metric reads
\begin{equation}
  d s^2 = -(1-\beta^2) d \eta^2 + 2 \beta d\eta\, d\chi + d\chi^2.
\end{equation}
We take $\eta$ as time, which means that $e_\eta$ is the evolution vector.
Its norm $e_\eta^2=-(1-\beta^2)$ may be positive,
negative or zero. Choosing $\beta^2=1$ yields an evolution along a
null direction.
However the tangent space to the constant time hypersurface
$\Sigma_{\eta_0}$ is
$T\Sigma_{\eta_0} =span\{ e_\chi|_{(\eta_0,\chi)}\}$
  and it is always spacelike since $e_\chi^2=1$.
It follows that the orthogonal direction is given by $e_t$ and only
the component along $e_t$ matters for the evolution, while
the remaining part is just a tangential shift associated with the
chosen coordinates.
The dynamics is therefore timelike.

In the same line of thought we can start from a lightcone metric $d s^2 = -2 d \xp\, d\xm + d x^2$ in a 3D Minkowski spacetime
and change variables as 
$\xp=u, \xm= v+ \beta u$
so that the new metric reads
\begin{equation}
  d s^2 = -2 \beta d u^2 - 2 d u\, d v + d x^2.
  \label{eq:normal_in_lc}
\end{equation}
We now take $u$ as time and $e_u$ as evolution vector.
It has norm $e_u^2=- 2\beta$ and may be also a timelike
vector if we choose $\beta>0$.
Nevertheless the constant time hypersurface $\Sigma_{u_0}$ has tangent space
$T\Sigma_{u_0} =span\{ e_v|_{(u_0, v, x)},  e_x|_{(u_0, v, x)}\}$.
This means that  it is always lightlike since $e_v^2=0$.
We can now compute the ``orthogonal'' vector.
Taking as image the usual Minkowski space
this is the null vector $N$ such that $g(N, e_v)=-1$
and $g(N, e_x)=0$, i.e. $N=e_u -\beta e_v = e_+$.
It is then only the component of $e_u$ along this null vector which
really determines the lightlike dynamics of this case.
Moreover this approach can yield a constant time hypersurface which is
neither lightlike nor spacelike but whose causal character
varies along the surface itself.
Consider again 3D Minkowski and this time change variables as 
$\xp=v + \oh \alpha u^2, \xm= u$ so that the metric becomes
\begin{equation}
  d s^2 = -2 \alpha u d u^2 - 2 d u\, d v + d x^2
  .
\end{equation}
We now take $v$ as time and the evolution vector $e_v$ is lightlike
since it has norm $e_v^2=0$.
The constant time hypersurface $\Sigma_{v_0}$ has tangent space
$T\Sigma_{v_0} =span\{ e_u|_{(u, v_0, x)},  e_x|_{(u, v_0, x)}\}$: since $e_u^2=-2\alpha u \lessgtr 0$, a piece of the surface is spacelike and another piece is timelike.
It doesn't therefore seem straightforward to define light-cone evolution relying only on 2).

Let us now look at the particle point of view.
In order to do so we consider
the following somewhat generic metric with a null
Killing vector $k=\partial_v$.
%
All metric
components are then $v$ independent and $g_{v v}=0$, so we can write\footnote{The $x$ dependence of the coefficients $h, f_i, l_i$ and $g_{ij}$ should in general be understood from any $x^i$ component.}
\begin{align}
  d s^2
  = &
  - 2 d u\, d v
  + h(u, x) d u^2
  + 2  f_i(u,x) d v d x^i
  + 2  l_i(u,x) d u d x^i \\ \nonumber
  & + g_{i j}(u, x) d x^i d x^j
  .
  \label{eq:generic_metric0}
\end{align}
The action for a massive particle in this background reads
\begin{align}
  S
  =&
  \int d\lambda
  \oh \left[
  \frac{1}{\ein}
  \left(
   h \dot u ^2
   + 2 l_i  \dot u\, \dot x^i
   + g_{i j} \dot x^i\, \dot x^j 
  - 2 \dot v\, \dot u
  + 2 f_i  \dot v\, \dot x^i
  \right)
  -
  {\ein}{m^2}
  \right]
  ,
\end{align}
where $\ein$ is the one dimensional einbein.

The first possible definition of light-cone corresponds to propagation
in the null direction $k_v$, i.e. to fix the diffeomorphisms using the
gauge
\begin{equation}
  v=\tau.
\end{equation}
Under this choice the action becomes
\begin{align}
  S_{g.f.}
  =&
  \int d\tau
  \oh \left[
  \frac{1}{\ein}
  \left(
   h \dot u ^2
   + 2 l_i  \dot u\, \dot x^i
   + g_{i j} \dot x^i\, \dot x^j 
  - 2 \dot u
  + 2 f_i  \dot x^i
  \right)
  -
  {\ein}\,{m^2}
  \right]
  .
\end{align}

It is then quite obvious that the map between velocities $(\dot u, \dot
x^i)$ and momenta $(p_u, p_i)$ is generically invertible.
It follows that the final Hamiltonian still depends on the
einbein $e$ and its elimination requires the use of a square root.
More precisely, the relation between velocities and momenta
\begin{equation}
  ( p_I)
  =
  \left( \begin{array}{c} p_u \\ p _i \end{array} \right)
  =
  \frac{1}{\ein} \left(\gamma_{I J } \dot x^J + \frac{A_I}{2} \right)
  =
  \frac{1}{\ein}
  \left( \begin{array}{c c} h & l_i \\ l _i & g_{i j} \end{array} \right)
  \left( \begin{array}{c} \dot u \\ \dot x^j \end{array} \right)
  +
  \frac{1}{2\ein}
  \left( \begin{array}{c} -2 \\ 2  f_i \end{array} \right)
\end{equation}
may be inverted when $\det \gamma \ne 0$ as effect of the gauge fixing,
but we have to get rid of the Lagrange multiplier $\ein$ which is still present. This can be done at the Lagrangian level or at the
Hamiltonian level.
In the latter case, the Hamiltonian with the Lagrange multiplier $\ein$ reads
\begin{equation}
  H_{g.f.}
  =
  \frac{\ein}{2} (\gamma^{ I J} p_I p_J + m^2)
  + \frac{1}{8 \ein} \gamma^{ I J} A_I A_J
 .
\end{equation}

Notice that $(\det \gamma\, d^{D-1}x^I) $ is precisely the volume of the constant time
hypersurface $\Sigma_{v_0}$ which is parameterized by the coordinates
$x^I=(u, x^i)$ whose induced metric is $d s^2|_\Sigma = \gamma_{I J} d x^I\,d x^J$.
Therefore a gauge fixed particle action with a square root is
equivalent to a constant time hypersurface which is not lightlike.
As noted before, this hypersurface may be locally spacelike or timelike. 

If we want a constant time null hypersurface it is better to fix the
gauge using $u$ so that $v$ parametrizes $\Sigma_{u_0}$ and the null vector
$e_v=\partial_v$ is tangent.
Nevertheless the previous condition is not sufficient.
In fact the induced metric can be written as
\begin{equation}
  d s^2|
  =
  -(E^{\underbar v} )^2 + \sum_i (E^{\underbar i} )^2
  =
  \left( \sqrt{ f_i \bar g^{i j} f_j} d v \right)^2
  + \sum_i
  \left[ \bar v^{\underbar i }{}_i
  \left( d x^i + \bar g^{i  j} f_j d v \right)
  \right]^2
  ,
\end{equation}
where
$\bar g_{i j} = g_{i j}$ is the metric $g$ restricted to indexes
$i,j,\dots$
and
$ \sum_{\underbar i}
\bar v^{\underbar i }{}_i \bar v^{\underbar i  }{}_j
  = \bar g_{i j}
$.
It then follows that
$ vol = E^{\underbar v} \wedge_{\underbar i} E^{\underbar i}
= \sqrt{f_i \bar g^{i j} f_j\, \det \bar g}
.
$
Then the constant time hypersurface $\Sigma_{u_0}$ is timelike
when $f_i\ne 0$ and spacelike when $f_i=0$.

Another way to reach the same conclusion is to discuss when
it is possible to choose the would-be light-cone gauge
\begin{equation}
  u=\tau
  .
\end{equation}
Upon this choice the gauge fixed action reads
\begin{align}
  S_{g.f.}
  =&
  \int d\tau
  \oh \left[
  \frac{1}{\ein}
  \left(
    -2 \dot v
   + g_{i j} \dot x^i\, \dot x^j 
   + 2 l_i   \dot x^i
   + h +2f_i \dot v
  \right)
  -
  {\ein}{m^2}
  \right]
  ,
\end{align}
which implies a non trivial e.o.m. for $v$:
\begin{equation}
\frac{d}{d \tau}
  \left\{
  \frac{1}{\ein}
  \left[\dot u
  -   f_i(u,x) \dot x^i
  \right]
    \right\}
    =
    0.
\end{equation}
We need therefore to take again $f_i(u,x)=0$ in order to have $p_v = -\frac{1}{\ein}$ and, as a consequence, an
Hamiltonian free of square roots.


Having understood the basic principles of the light-cone quantization, we can
consider another similar example.
It is noteworthy that this metric does not admit a null Killing vector
but still have a simple light-cone quantization.
We take
\begin{equation}
  d s^2
  =
  2 d r\, d u - f(u, r) d u^2 + h(u, r) d \theta^2 
,
\end{equation}
which is inspired to the 4D Vaidya metric where $h(u, r)=r^2$.
In this case we take as time $u$ so that the induced metric on the
constant time hypersurface $\Sigma_{u_0}$ is
\( d s^2 |_{\Sigma_{u_0}} = h(u_0, r) d \theta^2\)
which has null volume element $vol(\Sigma_{u_0}) =0$ and therefore
is lightlike. The spacetime volume element is instead
$vol= \sqrt{ h(u,r) } d u\, d r\, d \theta$ and can depend on the time
$u$. This does not happen in the real Vaidya metric.

As suggested above we fix the gauge $u=\tau$ and we read the gauge
fixed particle action to be
\begin{equation}
  S_{g.f.}
  =
  \int d \tau\, \oh \left\{
  \frac{1}{\zeta}
  \left[ 2 \dot r - f(\tau, r) + h(\tau, r) \dot \theta^2 \right]
  -
  \zeta m^2
  \right\}
  ,
\end{equation}
from which we derive the light-cone Hamiltonian
\begin{equation}
  H_{l.c.}
  =
  \frac{1}{2 p_r}
  \left[ \frac{p_\theta^2}{h(\tau, r)} + m^2 \right]
  +
  \oh f(\tau, r) p_r
  ,
\end{equation}
which is free of square roots and
where we have $p_r=\frac{1}{\zeta}$.
This possibility prompts for a question. Time-dependent backgrounds are usually associated with particle creation but the ground state in LCQFT is trivial and so there is no obvious alternative.
It could be that the solution is in the wave function as it has been proposed for the analogous problem in the spontaneous symmetry breaking case \cite{Rozowsky:2000gy}.

\subsection{A General Metric}
Therefore the metric we consider as examples in this paper is
\begin{align}
  d s^2
  =&
  - 2 d u\, d v
  + h(u, x) d u^2
  + 2  l_i(u,x) d u d x^i
  + g_{i j}(u, x) d x^i d x^j
  ,
  \label{eq:generic_metric}
\end{align}
which is the general pp wave metric.

The metric and its inverse read in matrix form as
\begin{align}
  g
  = &
  \parallel g_{\mu \nu} \parallel \, 
  =
  \left(
  \begin{array}{c c c}
    h & -1 & l_j
    \\
    -1 & 0 & 0
    \\
    l_i & 0 &  g_{i j}
  \end{array}
  \right)
  ,
  \nonumber\\
  g^{-1}
  = & 
  \parallel g^{\mu \nu} \parallel \, 
  =
  \left(
  \begin{array}{c c c}
    0 & -1 & 0
    \\
    -1 & -h + \bar l^{\,2} &  \bar l^j
    \\
    0 & \bar l^i & \bar g^{i j}
  \end{array}
  \right)
  ,
\end{align}
where $\bar g^{i j}$ is the inverse of $ g_{i j}$,
$\bar l^i = \bar g^{i j} l_j$,
and $\bar l^{\,2} = \bar g^{i j} l_i l_j$.
It happens however that $g^{i j} = \bar g^{i j}$, i.e. the restriction
to $i,j$ indexes of $g^{-1}$ matches $\bar g^{-1}$.

\subsubsection{Useful Expressions}
We can write the metric using the two choices of vielbein
\begin{align}
    d s^2
  =&
  - 2 d u\, \left[ d v + \oh (\bar l^{\,2}- h) d u \right]
  + g_{i j}(u, x) (d x^i + \bar l^i d u)
  (d x^j + \bar l^j d u)
  \nonumber\\
  =&
  - 2 d u\, \left( d v - \oh h d u - l_i d x^i \right)
  + g_{i j}(u, x) d x^i d x^j 
.
\end{align}
Especially the second one is well suited to perform the computations, so that
\begin{align}
  \left(\begin{array}{c}
    E^u \\ E^v \\ E^{\underline{i}}
  \end{array}
  \right)
  &=
  \left(\begin{array}{c c c}
    1 & 0 & 0\\
    -\oh h & 1 & l_j \\
    0 & 0 & \bar E^{\underline{i}}_j
  \end{array}
  \right)
  \left(\begin{array}{c}
    d u \\ d v \\ d x^j
  \end{array}
  \right)
  ,
  \nonumber\\
  \left(\begin{array}{c}
    d u \\ d v \\ d x^i
  \end{array}
  \right)
  &=
  \left(\begin{array}{c c c}
    1 & 0 & 0\\
    \oh h & 1 & l_{\underline{j}} \\
    0 & 0 & \bar E^i_{\underline{j}}
  \end{array}
  \right)
  \left(\begin{array}{c}
    E^u \\ E^v \\ E^{\underline{j}}
  \end{array}
  \right)
  ,
\end{align}
where $\bar E^{\underline{i}}_j$ is the vielbein w.r.t. the metric
$\bar g_{i j}(u, x)$ when $u$ is treated as a parameter.
From the previous expression we can immediately read the inverse
metric $g^{\mu\nu}$ and
the volume element to be
\begin{align}
  vol
  =&
  E^{\underline{u}}   E^{\underline{v}}   E^{\underline{2}}\dots   E^{\underline{D-1}}  
  =
  \sqrt{ |\bar g|} d u d v d ^{D-2} x
  ,
\end{align}
with $|\bar g |= \det (g_{i j})$.

It is straightforward  to derive the Christoffel symbols to be
\begin{alignat}{4}
 &\phantom{\Gamma^{u}_{\mu\nu }= 0}&&\,\,\,\,\,\Gamma^{u}_{\mu\nu }= 0,
  &&& \phantom{\Gamma^{u}_{\mu\nu }= 0}
  \nonumber\\
  &\Gamma^{v}_{u u }= - \oh \partial_u h 
  ,&&\,\,\,\,\, \Gamma^{v}_{u i } = - \oh \partial_i h
  ,&&&
  \Gamma^{v}_{i j } =\oh \partial_u g_{i j} -  \partial_{(i} l _{j)}
  ,
  \nonumber\\
  &\Gamma^{i}_{u u} =  \bar g^{i j}
  ( - \oh \partial_j h + \partial_u l_i )
  ,&&\,\,\,\,\,
  \Gamma^{i}_{u i }= \oh \bar g^{i l}
  ( \partial_u g_{l j} -2 \partial_{[ l} l_{j]} )
  ,&&&
  \Gamma^{i}_{j k }= \bar \Gamma^i_{j k}
  .
\end{alignat}
Finally the divergence of a current is
\begin{align}
  D^\mu J_\mu
  =\frac{1}{\sqrt{|\bar g|}}
  \Bigl\{
    &
    - \partial_u \left(\sqrt{|\bar g|}  J_v\right)
    \nonumber\\
    &
    + \partial_v \left[\sqrt{|\bar g|}
    \left( - J_u + (\bar l^{\,2} -h) J_v + \bar l^i J_i\right)\right]
    \nonumber\\
    &
    + \partial_i \left[\sqrt{|\bar g|}
    \left(  \bar l^i J_v + \bar g^{i j} J_j \right)\right] 
    \Bigr\}
  .
  \label{eq:NO_prel_divergence}
\end{align}





\subsection{Special Cases}

The previous metric includes the following special cases of which we
rapidly discuss the constant ``time'' hypersurfaces.

\begin{itemize}
\item
  The first one corresponds to
  Rosen coordinates shock wave family of metrics and reads 
\begin{equation}
  d s^2 = -2 d u\, d v+ g(u) (d z)^2 + \sum_{i=3}^D (d x^i)^2,
  \label{eq:metr:Ros}
\end{equation}
with $i=3,...D$.
Actually, we can distinguish different subcases:
\begin{align}
  \begin{cases}
    g(u)=1 \quad & \mbox{Minkowski}
    \\
    g(u)= (\Delta u)^2 \quad & \mbox{Null Boost Orbifold coordinates
      in Minkowski\footnotemark}
    \\
    g(u) = u^{2 A} \quad & \mbox{Light-cone Kasner-Rosen}
    \label{eq:metr:Kas-Ros}
\end{cases}
    ,
\end{align}
\footnotetext{See \cite{2002_seiberg,2020_arduino} for more details on this model.}
In particular $A=0,1$ correspond to Minkowski spacetime.
The expression \eqref{eq:metr:Kas-Ros} can also be generalized to the case with $P$ Kasner exponents $A_I$ (with $i=P+I\in\{P\!+\!1,...D\}$)
\begin{equation}
  d s^2= - 2 d u\, d v
  + \sum_{I=1}^{P} u^{2 A_I} (d z^I)^2
  + \sum_{i=P+1}^D (d x^i)^2
  .
  \label{eq:metr:genRos}
\end{equation}

The light-cone corresponds to take $u$ as evolution time.
In fact the induced metric on $\Sigma_{u_0}$ implies a vanishing
volume since $g_{v\, u}|_{u_0}=0$.
The evolution vector $\partial_u$ is null since the trajectories
$\gamma(\lambda; v_0, {\vz}_0, \vx_0)= (u=\lambda, v=v_0,
{\vz}={\vz}_0, \vx=\vx_0)$ are
lightlike and therefore they can be realized with physical observers.

Notice also for the implication on light-cone quantization
that the determinant of the metric is generically ($\sum_I
A_I \ne 0$) light-cone time-dependent and reads:
\begin{equation}
  \sqrt{- \det g}
  =
  \sqrt{ |g|}
  =
  |u|^{\sum_I A_I}
  .
\end{equation}
This means that the spacetime at $u=0$ is singular.
The singularity may be a coordinate singularity or a true singularity.
However the argument presented in the next section depends only on the fact that the volume element is light-cone time-dependent since we can always consider the evolution far from the singularity.


\item
The second class of family of metrics  can be
obtained from the previous one by changing to Brinkmann-Fermi coordinates\footnote{\label{foot:Rose_Brinkmann_change_coordinates}
The change of coordinates reads $v_R=v_B-\oh \sum_I A_I (x^I)^2/u$ and $z^I= x^I / |u|^{A_I}$ so that for $u_0\ne 0$ the constant time hypersurfaces are identical, 
i.e. $\Sigma_{u_0 (R)} = \Sigma_{u_0 (B)}$.}
and reads
\begin{equation}
  d s^2
  =
  - 2 d u\, d v
  +
   \frac{\sum_I A_I (A_I-1) (x^I)^2}{u^2} d u^2
  +
  \sum_I (d x^I)^2
  ,
\end{equation}
where, as before, for $A_I=0,1$ we get the Minkowski spacetime.

As before the light-cone propagation corresponds to choose $u$ as time.
This happens again because the induced volume element is zero.
However, differently from the previous case the evolution vector $\partial_u$ is
not lightlike since
the trajectories
$\gamma(\lambda; v_0, \vx_0)= (u=\lambda, v=v_0, \vx=\vx_0)$
have
$\dot \gamma^2= \frac{\sum_I A_I(A_I-1) (x_0^I)^2}{\lambda^2}$.
Therefore they can be interpreted as physical observers worldlines as
long as $\sum_I A_I(A_I-1) (x_0^I)^2 \le 0$.
The observer with $x^I_0=0$ is always physical and lightlike.

Finally we notice that in this case the determinant of the metric is trivial, i.e.
\begin{equation}
  \sqrt{- \det g}
  =
  \sqrt{ |g|}
  =1
  .\end{equation}

\end{itemize}


\section{Quantizing the Complex Scalar Density}
\label{sez:LC_scalar_generic}
We now consider the pp wave metric in order to show the first not
trivial point: the field to quantize is not the scalar field but a
scalar density.
Nevertheless this consideration is independent of the explicit metric taken
as example.
We will start by considering a simple particle model and then we
will treat the free complex scalar.

\subsection{The Particle Model}
\label{subsec:one_part_model}
In order to mimic the hallmark of light-cone approach, i.e. the appearance of only first order time derivatives, we consider the action
\begin{equation}
    S=\int d t 
    \left[f(t)(y \dot x - x \dot y) - h(x,y,t)\right],
    \label{eq:act1}
\end{equation}
from which we derive the following equations of motion:
\begin{equation}
    \dot x = \frac{1}{2 f(t)}\frac{\partial h}{\partial y} -\frac{\dot f(t)}{2 f(t)}x, \qquad
    \dot y = -\frac{1}{2 f(t)}\frac{\partial h}{\partial x} -\frac{\dot f(t)}{2 f(t)}y.
    \label{eq:eom1}
\end{equation}
If we proceed in a naive way and apply the Dirac procedure for
constrained systems we get the classical Dirac bracket
\begin{equation}
\{ x, y \}_{D B}= \frac{1}{2 f(t)}
.
\end{equation}  
The same result can be obtained if we read from the action the
symplectic form $\omega= f(t) ( y d x - x d y )$.
Unfortunately these approaches are flawed.
The reason is that a symplectic form cannot depend on other
coordinates than the symplectic coordinates, i.e. the explicit time
dependence is not allowed.
This fault can be seen directly by checking that there is no
Hamiltonian $H(x, y, t)$ which gives the previous equations of motion using
the Dirac bracket.
In facts from $\dot x = \{x, H\}_{DB}$ we get $H=h- \dot f x y$
while from $\dot y = \{y, H\}_{DB}$ we get $H=h + \dot f x y$.
This means that we cannot take $x$ and $y$ as coordinates of the phase space.

Now we would like to give an Hamiltonian interpretation to these \eom{}s
but, as the unusual form of \eqref{eq:eom1} and the previous
discussion suggest, we cannot treat $x$ and $y$ as canonical
variables.
We are indeed forced to perform a change of variables at the Lagrangian level.\\
There are many ways of redefining the coordinates and we take for example
(any other redefinition which eliminates the $t$ factor will do):
\begin{equation}
    x=\frac{1}{\sqrt{2f(t)}}\hat{x}, \qquad y=\frac{1}{\sqrt{2f(t)}}\hat{y}.
    \label{eq:part_redef}
\end{equation}
The action \eqref{eq:act1} now reads
\begin{equation}
     S=\int d t 
    \left[\frac{1}{2}(\hat{y} \dot{\hat{x}} - \hat{x} \dot{\hat{y}}) - h\left(\frac{1}{\sqrt{2f(t)}}\hat{x},\frac{1}{\sqrt{2f(t)}}\hat{y},t\right)\right],
      \label{eq:act2}
\end{equation}
from which follow the usual \eom{}s
\begin{equation}
    \dot{\hat{x}} = \frac{\partial h}{\partial \hat{y}}, \qquad
    \dot{\hat{y}} = -\frac{\partial h}{\partial \hat{x}},
    \label{eq:eom2}
\end{equation}
and therefore the identifications
\begin{equation}
    q\equiv\hat{x}, \quad p\equiv\hat{y}, \quad H(\hat{x},\hat{y},t)\equiv h\left(\frac{1}{\sqrt{2f(t)}}\hat{x},\frac{1}{\sqrt{2f(t)}}\hat{y},t\right).
    \label{eq:phase_space}
\end{equation}
Given this well defined phase space
our original variables can be seen as ``composite
operators'' of the true canonical variables and 
we can compute the Poisson bracket of the original variables from this
point of view and get
\begin{equation}
    \{x,y\}=\frac{1}{2f(t)}.
\end{equation}
We can also recover the \eom{}s \eqref{eq:eom1}
if we take into account the explicit dependence of $x$ and $y$ on $t$, i.e.:
\begin{equation}
    \dot x = \{ x, H \} + \frac{\partial x}{\partial t}, \qquad \dot y = \{ y, H \} + \frac{\partial y}{\partial t}.
\end{equation}
In other words this means that $x$ and $y$ have to be seen as time
dependent functions defined on the phase space
\eqref{eq:phase_space}. 
The bottom line of this discussion is that the variables redefinitions
are necessary to get rid of the time dependence
(and also of any additional constant) which appears in front of the
"kinetic" terms of \eqref{eq:act1}.
We will see in a moment how to rephrase this in field theory. \\

\subsection{The Complex Scalar Field}
We begin with the classical treatment and then we move to the quantum one.
This background has essentially
been considered before in the usual formalism in \cite{Gibbons:1975jb}.
The starting point is the kinetic part of the action which reads:
\begin{align}
  S_2\!
  = \!&
  \int \!\!d u\, d v\, d^{D-2} x
  \sqrt{|\bar g|}
  \Bigl\{
  \partial_u \phi^*\,  \partial_v \phi
  +
  \partial_v \phi^*\,  \partial_u \phi
  +
  \left[  h(u,x) - \bar l^{\,2}(u,x) \right] \partial_v \phi^*\,  \partial_v \phi
  \nonumber\\
  &
  - \bar l^i(u,x) ( \partial_i \phi^*\,  \partial_v \phi
  +
  \partial_v \phi^*\,  \partial_i \phi )
  -
  g^{i j}(u,x)  \partial_i \phi^*\,  \partial_j \phi
  -M^2 \phi^* \phi
  \Bigr\}
  \nonumber\\
  =\!&
  \int \!\!d u\, d v\, d^{D-2} x
  \sqrt{|\bar g|}
  \Bigl\{
  \partial_u \phi^*\,  \partial_v \phi
  +
  \partial_v \phi^*\,  \partial_u \phi
  +
  h(u,x)  \partial_v \phi^*\,  \partial_v \phi
  \nonumber\\
  &
  -
  g^{i j}(u,x)
  \left[ \partial_i \phi^*\, + l_i(u,x) \partial_v \phi^* \right]
    \left[ \partial_j \phi + l_j(u,x) \partial_v \phi \right]
  -M^2 \phi^* \phi
  \Bigr\}
  ,
\end{align}
where $\bar g_{i j} = g_{i j}$ is the pullback metric on $u, v$
constant and as before
$\bar l^i = \bar g ^{i j} l_j$,
$\bar l^{\,2} = \bar g ^{i j} l_j l_j$ and 
$|\bar g|=\det( g_{i j} )$.

In the following  we consider the
complex scalar field coupled to an electromagnetic background.
In order to simplify the computations we take $A_{\mu}$
invariant under the same Killing vector $e_v=\partial_v$.
Moreover, we choose the light-cone
gauge
\begin{equation}
  A_v=0,~~~~
  A_u=A_u(u,x)~~~~
  A_i=A_i(u,x)
  \label{eq:Killing_inv_LC_gauge_em_bcg}
  ,
\end{equation}
so that the action reads:
\begin{align}
  S_2
  =&
  \int d u\, d v\, d^{D-2} x
  \sqrt{|\bar g(u,x)|}
  \Bigl\{
  \cD_u \phi^*\,  \partial_v \phi
  +
  \partial_v \phi^*\,  \cD_u \phi
  +
  h(u,x)  \partial_v \phi^*\,  \partial_v \phi
  \nonumber\\
  &
  -
  g^{i j}(u,x)
  \left[ \cD_i \phi^*\, + l_i(u,x) \partial_v \phi^* \right]
    \left[ \cD_j \phi + l_j(u,x) \partial_v \phi \right]
  -M^2 \phi^* \phi
  \Bigr\}
  ,
\end{align}
where $\cD_\mu = \partial_\mu -i e\, A_\mu$.

As shown in the particle model in the previous section,
since $\sqrt{|\bar g(u,x)|}$ does depend on time $u$ we cannot
proceed as usual not even using the more formal approach of the Dirac brackets. We need to
redefine the fields and the minimal field redefinition for the
light-cone quantization is 
\begin{align}
  \phi(u, v, x) = |\bar g(u, x)|^{-\frac{1}{4}} \hphi(u, v, x)
  ,
\end{align}
where the new field in not anymore a scalar but a scalar
density\footnote{ This scalar density is the object that makes the integrability condition trivial in \cite{hayward1993}, in particular after the null reduction as in \cite{Sachs:2021mcu}. }.
This rescaling is obviously necessary for all the other spins and does not change the physics.\\
The action becomes:
\begin{align}
  S_2
  =&
  \int d u\, d v\, d^{D-2} x
  \Bigg[
   \cD_u \hphi^*\,  \partial_v \hphi
  +
  \partial_v \hphi^*\,  \cD_u \hphi
  +
  h\,  \partial_v \hphi^*\,  \partial_v \hphi
  \nonumber\\
  &
  -
  g^{i j}\,
  \left(
  |\bar g|^{\frac{1}{4}} \cD_i \frac{\hphi}{|\bar g|^{\frac{1}{4}}}
  + l_i \partial_v \hphi
  \right)^*\,
  \left(
  |\bar g|^{\frac{1}{4}} \cD_j \frac{\hphi}{|\bar g|^{\frac{1}{4}}}
  + l_j \partial_v \hphi
  \right)
  -M^2 \hphi^* \hphi
  \Bigg]
  .
\end{align}
We perform now a Fourier transform w.r.t. $v$ as
\begin{align}
  \hphi(u, v, x)
  =&
     \int_{-\infty}^\infty \frac{d k_v}{(2\pi)^{1/2}}
     e^{i k_v v}
     \frac{1}{\sqrt{2 |k_v|}} \thphi(u, k_v, x).
     %
     %
     %
     %
\end{align}
The previous expression can be written for the original scalar field
as
\begin{align}
  \phi(u, v, x)
  =
  &
  \frac{ 1 }{ |\bar g(u, x)|^{\frac{1}{4}} }
  \int_{-\infty}^0 \frac{d k_v}{(2\pi)^{1/2}}
  e^{i k_v v}
  \frac{1}{\sqrt{2 |k_v|}} \thphi(u, k_v, x)
  \nonumber\\
  &
  \phantom{   \phi(u, v, x)
  =
  }
  +
  \frac{ 1 }{ |\bar g(u, x)|^{\frac{1}{4}} }
  \int_{-\infty}^0 \frac{d l_v}{(2\pi)^{1/2}}
  e^{- i l_v v}
  \frac{1}{\sqrt{2 |l_v|}} \thphi(u, -l_v, x)
  ,
\label{eq:LCQFT_Fourier_phi}
\end{align}
in a way which is useful to compare with the second
quantization of the particle.
In particular it can be interpreted as the sum of two particles,
the first one with wave function $\thphi(u, k_v, x)$ ($\kv<0$) and
the second one with wave function $\thphi(u, -k_v, x)$
($\kv<0$)\footnote
{Notice that with our notation
  $\widetilde{\hat\phi^*(v)}(k_v)= \left( \thphi(-k_v) \right)^*$ so
  that we can express $\thphi(-\kv)$ using the field
  $\widetilde{\hat\phi^*(v)}(k_v)$ which has the natural range to be
  interpreted as a particle.
  }.
This happens because we are quantizing the complex scalar;
if we chose instead a real scalar $\phi \rightarrow \frac{1}{\sqrt{2}}\phi_\R$,
we would get $\thphi_\R (u, -k_v, x) = \thphi_\R(u, k_v, x)^*$ and
therefore only one particle with wave function $\thphi_\R (u, k_v, x)$.

The action can then be written in a form which can be interpreted
as the sum of two actions for two particles:
\begin{align}
  S_2
  =&
  \int d u\, d^{D-2} x \int_{-\infty}^0 d \kv
  \Biggr\{ \Biggl[
    i (\thphi(\kv))^*\,  \partial_u \thphi(\kv) 
  \nonumber\\
  &\!\!\!\!\!\!\!\!\!
  +
  \frac{1}{2 \kv} g^{i j}\,
  \left(
  |\bar g|^{\frac{1}{4}} \cD_i \frac{ \thphi(\kv) }{ |\bar g|^{\frac{1}{4}} }
  + i \kv l_i\, \thphi(\kv)
  \right)^*
  \left(
  |\bar g|^{\frac{1}{4}} \cD_j \frac{ \thphi(\kv) }{ |\bar g|^{\frac{1}{4}} }
  + i \kv l_j\, \thphi(\kv)
  \right)
  \nonumber\\
  &\!\!\!\!\!\!\!\!\!
  +\left(
  - e  A_u
  -\oh \kv h\, 
  + \frac{M^2}{2 \kv}
  \right) (\thphi(\kv))^* \thphi(\kv)
  \Biggr]
  \nonumber\\
  &
  \phantom{
  \int d u\, d^{D-2} x \int_{-\infty}^0 d \kv
  \Biggr\{
  } 
  \!\!\!\!\!+
  \Biggl[
    i (\thsphi(\kv))^*\,  \partial_u \thsphi(\kv) 
  \nonumber\\
  &\!\!\!\!\!\!\!\!\!
  +
  \frac{1}{2 \kv} g^{i j}\,
  \left(
  |\bar g|^{\frac{1}{4}} \cD_i^* \frac{ \thsphi(\kv) }{ |\bar g|^{\frac{1}{4}} }
  - i \kv l_i\, \thsphi(\kv)
  \right)^*
  \left(
  |\bar g|^{\frac{1}{4}} \cD_j^* \frac{ \thsphi(\kv) }{ |\bar g|^{\frac{1}{4}} }
  - i \kv l_j\, \thsphi(\kv)
  \right)
  \nonumber\\
  &\!\!\!\!\!\!\!\!\!
  +\left(
  + e  A_u
  + \oh \kv h\, 
  +\frac{M^2}{2 \kv}
  \right) (\thsphi(\kv))^* \thsphi(\kv)
  \Biggr]
  \Biggr\}
  ,
  &
  \label{eq:S_2nd_quant_LCFT_gen_bck}
\end{align}
where we have integrated by parts in time $u$ in order to get a
canonical $p \dot q$
and we have dropped the boundary term
$ \int d v \partial_v(\dots \hphi^* \hphi)$ under the assumption that
$\hphi \rightarrow 0$ as $v \rightarrow \pm \infty$.
This asymptotic behavior is also important for getting a conserved
charge as discussed below.

Notice that the interpretation as sum of two independent particles is
possible because each contribution in square brackets is real (up to
boundary terms).
Moreover, only when using the natural fields
$\thphi(\kv)$ and $\thsphi(\kv)$ ($\kv<0$) it appears clearly also in
the covariant derivative $\cD_i^*$ that the
antiparticle described by $\thsphi(\kv)$ has the opposite charge $-e$.

In the real case the two contributions are equal so that the action
for a real scalar reads:
\begin{align}
  & S_{2,\, real}
  =
  \int d u\, d^{D-2} x \int_{-\infty}^0 d \kv
  \Biggl[
  i (\thphi_\R(\kv))^*\,  \partial_u \thphi_\R(\kv)
  \nonumber\\
  &
  +
  \frac{1}{2 \kv} g^{i j}\,
  \left(
  |\bar g|^{\frac{1}{4}} \partial_i \frac{ \thphi_\R(\kv) }{ |\bar g|^{\frac{1}{4}} }
  + i \kv l_i\, \thphi_\R(\kv)
  \right)^*
  \left(
  |\bar g|^{\frac{1}{4}} \partial_j \frac{ \thphi_\R(\kv) }{ |\bar g|^{\frac{1}{4}} }
  + i \kv l_j\, \thphi_\R(\kv)
  \right)
  \Biggr]
  .
  &
  \label{eq:S_2nd_quant_LCFT_gen_bck_real_case}
\end{align}

The canonical coordinates are
$q\sim \thphi(\kv, x),~\left( \thphi(-k_v, y) \right)^*$
and $p\sim i \left( \thphi(k_v, y) \right)^*, ~i \thphi(-k_v, y)$, with no time dependence since we are in Hamiltonian formalism,
and the canonical commutation relations are (for $\kv <0$):
\begin{align}
  [  \thphi(\kvN 1, x_1) ,~ \thphi(\kvN 2, x_2)^* ]
  =\,\, &
  \delta(\kvN 1- \kvN 2)\, \theta(-\kv_1)\,
  \delta^{D-2}(x_1 - x_2)
  ,
  \nonumber\\
    [  \thsphi(\kvN 1, x_1) ,~ \thsphi(\kvN 2, x_2)^* ]
  =\,\, &
  \delta(\kvN 1- \kvN 2)\, \theta(-\kv_1)\,
  \delta^{D-2}(x_1 - x_2)
  .
\label{eq:LCQFT_can_brackets}
\end{align}

As a consequence, if we would consider the Schrödinger formalism the
wave functional would depend on the scalar density and not on the
field, i.e. \(\Psi( \hat \phi(v, x), u) \).

\subsection{The Light-Cone Field Expansion and Quantization}
\label{sec:LCQFT_field_expandion}
We now want to expand the fields in the Heisenberg representation in modes
to read the creation and annihilation operators.
Looking at \eqref{eq:LCQFT_can_brackets} it seems natural to treat 
$\thphi(u, \kv, x)$ and $\thsphi(u, \kv, x)$ separately and then to join
their expansions using \eqref{eq:LCQFT_Fourier_phi}.
The \eom\, for $\thphi(u, \kv, x)$ ($\kv<0$)
is like a Schrödinger equation and reads:
\begin{align}
  i \partial_u \thphi(u, \kv, x)
  =\, &
  \frac{1}{2 \kv}
  \left[ - \nabla_i g^{i j} \nabla_j + M^2 \right] \thphi(u, \kv, x) \nonumber \\ 
  & - \left[ e A_u - \oh \kv h \right]\thphi(u, \kv, x)
  .
\end{align}
We can then introduce an orthonormal complete basis
$\{ \tPsi_{(n,\pv)}(u, \kv, x ; e) \}$.
These functions are orthonormal \wrt\, the ``spacial coordinates'' $\kv, x$,
i.e. for all times $u$ we have
\begin{equation}
  \int^0_{-\infty} d\kv \int d^{D-2} x\,
  \tPsi_{(m,\pv)}^*(u, \kv, x ; e)\, \tPsi_{(n,q_v)}(u, \kv, x ; e)
  = \delta_{m,n} \delta(\pv - q_v)
  .
\end{equation}
Notice that we have explicitly shown the dependence on the charge $e$. 
This basis can be obtained from the time evolution of
the orthonormal complete basis $\{ \tpsi_n(u_0, x ; \kv, e) \}$
as\footnote{
  The $\delta(\kv -\pv)$ factor may at first sight appear strange but it is nothing more than
  the wave function of the free particle in momentum space associated with the Hamiltonian $H=\frac{p^2}{2m}$.
  }
\begin{equation}
  \tPsi_{(n,\pv)}(u, \kv, x ; e)
  =
  \tpsi_n(u_0, x ; \kv, e)\, \delta(\kv -\pv)
  .
\end{equation}
The orthonormal complete basis $\{ \tpsi_n(u_0, x ; \kv, e) \}$ is
associated with the stationary Schrödinger equation
\begin{align}
  & \left[
  \frac{1}{2 \kv}
  \left( - \nabla_i g^{i j} \nabla_j + M^2 \right) 
  - \left( e A_u + \oh \kv h \right)
  \right]_{u=u_0}   \tpsi_n(u_0, x; \kv, e) \nonumber \\
 & = E_n   \tpsi_n(u_0, x; \kv, e)
  ,
\end{align}
where $\kv$ is considered a parameter so that 
\begin{equation}
  \int d^{D-2} x\,
  \tpsi_m^*(u_0, x ; \kv, e)\, \tpsi_n(u_0, x ; \kv, e)
  = \delta_{m,n}
  .
\end{equation}
Then we can expand the field $\thphi_H(u, \kv, x)$ for $\kv<0$
in the Heisenberg picture as:
\begin{align}
  \thphi_H (u, \kv, x)
 & =
  \sum_n  \int^0_{-\infty} d\pv a_{(n, \pv) H}(u)\, \tPsi_{(n,\pv)}(u, \kv, x ; e) \nonumber \\
 & =
  \sum_n  a_{(n, \kv) H}(u)\, \tpsi_{n}(u, x; \kv, e)
  ,
\end{align}
where the operators $a_{(n, \kv) H}(u)$ are actually constant because of the \eom.
In a similar way we can expand:
$\thsphi_H(u, \kv, x)$ for $\kv<0$
\begin{equation}
  \thsphi_H (u, \kv, x)
  =
  \sum_n b_{(n, \kv) H}(u)\, \tpsi_n^*(u, x; \kv, -e)
  ,
\end{equation}
where the operators $b_{(n, \kv) H}(u)$ are again constant because of the \eom.
Since we know from \eqref{eq:LCQFT_can_brackets} that the basis $\{ \tPsi_{(n,\pv)}(u, \kv, x ; e) \}$ is orthonormal, we
get the usual commutation relation:
\begin{equation}
  [ a_{(m,\kv) H},\, a_{(n, \pv) H}^\dagger ]
  =
  [ b_{(m, \kv) H},\, b_{(n, \pv) H}^\dagger ]
  = \delta_{m,n} \delta(\kv-\pv)
  .
\end{equation}
The light-cone vacuum is defined also as
\begin{equation}
     a_{(m,\kv)} |{\Omega}\rangle
= 
b_{(m,\kv)} |{\Omega}\rangle
=0
.
\end{equation}
Finally, we can expand the original field in the Heisenberg picture as
\begin{align}
  & \phi_H(u, v, x)
  = \,
  \frac{ 1 }{ |\bar g(u, x)|^{\frac{1}{4}} }
  \int_{-\infty}^0 \frac{d k_v}{(2\pi)^{1/2}}
  \frac{1}{\sqrt{2 |k_v|}}
  \nonumber\\
  &
  \left[
    e^{i k_v v}
    \sum_n a_{(n, \kv) H}\, \tpsi_n(u, x; \kv, e) +
    e^{-i k_v v}
    \sum_n b_{(n, \kv) H}^\dagger\, \tpsi_n^*(u, x; \kv, -e)
    \right].
  \end{align}
  
  Differently from the usual second order evolution, the creator and annihilation operators can be obtained without time derivatives as
\begin{alignat}{3}
    a_{(m, l_v)}
    &=
    \int d^{D-2} x \frac{d v}{ \sqrt{ 2\pi} }
    e^{-i l_v v} 
    |\bar g(u, x)|^{\frac{1}{4}} 
    \tpsi_m^*(u, x; l_v, e)
    \phi_H(u, v, x)
    ,~~  && l_v>0
    ,
    \nonumber\\
        b^\dagger_{(m, -l_v)}
    &=
    \int d^{D-2} x \frac{d v}{ \sqrt{ 2\pi} }
    e^{-i l_v v} 
    |\bar g(u, x)|^{\frac{1}{4}} 
    \tpsi_n(u, x;  -l_v, -e)
    \phi_H(u, v, x)
    ,~~  && l_v<0
    .
\end{alignat}
As an application for the special cases considered it follows that the vacua for the Rosen and Brinkmann coordinates are the same.
This happens since the equal time hypersurfaces are the same and $v_R = v_B + \dots $ (as described in footnote \ref{foot:Rose_Brinkmann_change_coordinates}), where $\dots$ are terms independent from $v$.
It then follows $k_{v\, R}= k_{v\, B}$ so that $a_R$ can be expressed using $a_B$ only and therefore the two vacua are the same.

\section{Second Quantization of the Particle}
In this section we would like to explore how the second quantization
of the particle on the light-cone is connected to the light-cone
quantization of the scalar field.

\subsection{The Particle Action}

The action for the particle in a generic gravitational and
electromagnetic background reads:
\begin{align}
  S_{particle}
  &=\int d\lambda\,
    \left(
    -m
    \sqrt{
    -g_{\mu\nu}(x) \frac{d x^\mu}{ d \lambda}
    \frac{d x^\nu}{ d\lambda}
    }
    +\eph
    A_\mu(x)
    \frac{d x^\mu}{ d \lambda}
    \right)
    \nonumber\\
  &
    =\int d\lambda\,
    \ein(\lambda)
    \left[
    \oh
    \left(
    g_{\mu\nu}(x) \frac{d x^\mu}{ \ein(\lambda) d \lambda}
    \frac{d x^\nu}{ \ein(\lambda) d\lambda}
    - m^2
    \right)
    +\eph
    A_\mu(x)
    \frac{d x^\mu}{ \ein(\lambda) d \lambda}
    \right]
    ,
\end{align}
where $\eph$ is the physical electric charge and
$d s^2_{world-line}
= -\ein^2 (d\lambda)^2$
is the worldline metric.
Notice that in order to reproduce the original action we need
\begin{equation}
  \ein >0
  ,
\end{equation}
and this constraint is important in the following when considering the range of
the $v$ momentum.
The action has the diffeomorphism invariance
\begin{equation}
  d\lambda\, \ein(\lambda) =     d\tau\, \ein'(\tau),~~~~
  x^\mu(\lambda) = x'^{ \mu}(\tau)
  .
\end{equation}
The e.o.m read
\begin{align}
  - \ein^2 \frac{\delta S}{\delta \ein}
  = \, &
    \frac{\dot x^2}{\ein^2 }+m^2
  =0,
     \nonumber\\
  \frac{\delta S}{\delta x^\mu}
  =&
  -\frac{d}{ d\lambda}\left(
  g_{\mu\nu}(x) \frac{d x^\nu}{ \ein(\lambda) d \lambda}
  \right)
  +\eph F_{\mu\nu}(x) \frac{d x^\nu}{ d \lambda}
  =0.
\end{align}

\subsection{Particle in pp Wave Metric in Light-Cone Gauge}
Consider now the previous action
in the  metric \eqref{eq:generic_metric}
coupled to the electromagnetic background
\eqref{eq:Killing_inv_LC_gauge_em_bcg}, which as we know are invariant under
the same null Killing vector.
We gauge fix the diffeomorphisms as $u=\tau$ so that the action becomes
\begin{align}
  S_{l.c.}
  =&
  \int d \tau
  \Biggl\{
  \frac{1}{\ein}
  \left[
     - \dot v
    + \oh h(\tau, x)
    + \oh g_{i j}(\tau, x) \dot x^i \dot x^j
    +  l_i(\tau,x) \dot x^i
    \right]
  \nonumber\\
  &
  \phantom{  \int d \tau  \Biggl\{ }
  + \eph\,A_u(\tau, x)
  + \eph\,A_i(\tau,x) \dot x^i
  - \oh \zeta m^2
  \Biggr\}
  \nonumber\\
  =&
  \int d \tau
  \Biggl\{
    + \pv \dot v
    + p_i \dot x^i
    \nonumber\\
    &
  \phantom{  \int d \tau   \Biggl\{ }
    -\Bigl[
      - \frac{1}{2 \pv} \bar g^{i j}
      \left( p_i +  l_i p_v  - \eph\,A_i \right)
      \left( p_j +  l_j p_v  - \eph\,A_j \right)
      \nonumber\\
    &
  \phantom{  \int d \tau  \Biggl\{ - \Bigl[}      
      - \frac{m^2}{2 \pv}
      + \eph\,A_u
      + \oh h p_v
      \Bigr]
    \Biggr\}
    ,
\end{align}
where $\bar g^{i j}$ is the inverse of the metric
$\bar g_{i j} = g_{i  j}$ and not of $g_{\mu \nu}$.  
In this formulation $(h, l_i)$ acts as a kind of supplementary gauge field.

\subsubsection{Light-cone Hamiltonian Formalism and Quantization}
We can read the Poisson brackets
\begin{equation}
  \{ v, p_v \} = \{ x^i, p_i \} = 1
  ,
\end{equation}
and the classical light-cone Hamiltonian
\begin{align}
  H_{l c (classical)}(\tau, p_v, x^i, p_i)
  =
  &
  - \frac{1}{2 \pv} \bar g^{i j}
  \left( p_i +  l_i p_v  - \eph\,A_i \right)
  \left( p_j +  l_j p_v  - \eph\,A_j \right)
  \nonumber\\
&
- \frac{m^2}{2 \pv}
  + \eph\,A_u
  + \oh h p_v
.
\end{align}
Now this Hamiltonian suffers from ordering problems.
We want an Hermitian Hamiltonian but this is not uniquely fixed since
if we change the measure of integration we get different Hamiltonians:
$H= p_i \bar g^{i j} p_j$ is Hermitian \wrt\,
$vol= d^{D-2} x$,
while 
$H= \frac{1}{ \sqrt{|\bar g| }}p_i \sqrt{|\bar g|} \bar g^{i j} p_j$
is Hermitian \wrt\, $vol= \sqrt{|\bar g|} d^{D-2} x$.
Moreover, even when we fix the volume element we do not get a unique
result.
Indeed, let's consider a light-cone Hamiltonian which is
hermitian \wrt\, to $vol=\mu(\tau, x) d^{D-2} x\, d\pv$
\footnote{The hemiticity \wrt\, $d\pv$ is trivial but in the measure
  we need it since $p_v$ appears on the same level of $x^i$.}
and reduces to
the classical one but differs quantum mechanically: it
can be written as
\begin{align}
  H_{lc (1st), \mu,  \rho, \sigma}(x, \pv)
  =\, & + \frac{1}{2 \pv}
  \frac{1}{ \mu(\tau,x) \rho(\tau,x) }
  \nabla_i
  \left(   g^{i j}(\tau,x) \sigma(\tau,x)
  \nabla_j
  \frac{1}{ \rho(\tau,x) }
  \right) \nonumber \\
  & +
  V(\tau, x)
,
  \label{eq:gen_H_1st_quant_particle_gen_bck}
\end{align}
where $\rho(\tau,x)$ and $\sigma(\tau,x)$ are arbitrary functions and
we have introduced the ``gauge'' covariant derivative
\begin{equation}
  \nabla_j
  =
  i p_j  - i \left(  \eph\,A_i  -  l_j p_v \right)
.
\end{equation}
If we want to reproduce the rescaled complex scalar Hamiltonian
we must set
\begin{equation}
  \mu=1,~~~~
  \sigma= \sqrt{|\bar g|},~~~~
  \rho=\sqrt[4]{|\bar g|},
  \label{eq:mu_rho_sigma_scalar}
\end{equation}
i.e. we need an Hamiltonian Hermitian \wrt\, $vol= d^{D-2} x$ and we need to know that we are considering a scalar density of weight $\frac{1}{4}$.
This information has to be supplied and it does not come out of the
formalism automatically.
Finally the first quantized Hermitian quantum Hamiltonian
which reproduces the scalar action under a second quantization can be written as
\begin{align}
  H_{l c (1st)}
  =&
  + \frac{1}{2 \pv}
  \frac{1}{ | \bar g| ^{\frac{1}{4} } }
  \nabla_i
  \left( \sqrt{|\bar g|}  g^{i j}
  \nabla_j
  \frac{1}{ | \bar g|^{\frac{1}{4} } }
  \right)
  + \eph\,A_u
  + \oh h p_v
  .
  \label{eq: H_LC_1st}
\end{align}
This expression follows from the  usual one
$\frac{1}{ \sqrt{|\bar g| }}p_i \sqrt{|\bar g|} \bar g^{i j} p_j$,
which is Hermitian \wrt\,
$ vol= \sqrt{|\bar g|} d^{D-2} x$, by replacing
$\nabla_i \rightarrow
{| \bar g| }^{\frac{1}{4} } 
\nabla_i \frac{1}{ {| \bar g| }^{\frac{1}{4} } }$
as suggested by the replacement of a scalar with a scalar density.

In the case of the pp wave metric in Rosen coordinates, and for all
the other metrics whose metric determinant depends only on light-cone
time, the naive connection works since $\rho(\tau)$ filters trough the
spacial derivatives.

\subsubsection{Second Quantization of the Particle}

Since we are dealing with a time-dependent Hamiltonian there is no
energy conservation and therefore we cannot find a
basis of energy eigenfunctions.
We can proceed as done in section \ref{sec:LCQFT_field_expandion}.
We consider the instantaneous Hamiltonian $H_{l c (1st)}(\tau_0)$ which is Hermitian
and therefore we can
find an instantaneous basis
$\{ \tPsi_a(\tau_0, \pv, x) = \tpsi_n(\tau_0, x, \kv) \, \delta(\pv-\kv) \}$
whose elements are labeled by $a=(n, \kv)$.
Using this basis we can expand the second quantized field in
the Heisenberg picture
(which will become the Interaction picture
after the introduction of interactions) as
\begin{align}
  \tPsi_H(\tau, \pv, x)
  \!=&
  \sum_a\negthickspace\negthickspace\negthickspace\negthickspace\negthickspace
  \int A_{a H}(\tau; \tau_0)\,  \tPsi_a(\tau_0, \pv, x)
=\!
  \sum_n\! \hat A_{(n, \pv) H}(\tau; \tau_0)\,  \tpsi_n(\tau, \pv, x; \tau_0)
  ,
\end{align}
where $A_{a H}(\tau; \tau_0)$ are labeled by
$\tau_0$ but also by $a$ which includes $\kv$.
The $A_{a H}(\tau; \tau_0)$ are annihilators of the second
quantized vacuum $|\Omega\rangle$
\begin{equation}
  A_{a H}(\tau; \tau_0) |\Omega\rangle
  =
  0
\end{equation}
and satisfy the harmonic oscillator algebra
\begin{align}
  [ A_{a H}(\tau; \tau_0), A_{b H}^\dagger(\tau; \tau_0)  ]
  =
  \delta_{a, b}
  = \delta(n_a-n_b)\,\delta(k_{v\, a} - k_{v\, b})
  .
\end{align}
While the previous two equations are kinematical statements
for $\tau=\tau_0$, for all the other possible $\tau$ they are
dynamical  and follow from the second quantized action:
\begin{align}
  S_{2 (2nd)}
  \!=&
  \!\!\int \!d\tau\, d^{D-2}x\,
  \!\!\int^0_{-\infty}\!\!\!\!\! d\pv\,
  \tPsi_H^\dagger(\tau, \pv, x)
  \left[ i \partial_\tau - H_{l c (1st)}(\tau, x, \pv, p) \right]
  \tPsi_H(\tau, \pv, x)
  \nonumber\\
  =&
  \!\!\int\! d\tau\, d^{D-2}x\,
  \int^0_{-\infty}\!\!\!\!\! d\pv\,
  \tPsi_H^\dagger(\tau, \pv, x) \Biggl\{ i \partial_\tau
    -
    \Biggl[
    \frac{1}{2 \pv}
    \frac{1}{ | \bar g| ^{\frac{1}{4} } }
    \nabla_i
    \left( \sqrt{|\bar g|}  g^{i j}
    \nabla_j
    \frac{1}{ | \bar g|^{\frac{1}{4} } }
    \right)
    \nonumber\\
    &
    - \frac{m^2}{2 \pv}
    + \eph A_u
    + \oh h p_v
    \Biggr]
    \Biggr\}
  \tPsi_H(\tau, \pv, x)
  %
  \\
  =&
  \sum_{a, b}\negthickspace\negthickspace\negthickspace\negthickspace\negthickspace\int
  A_{a H}^\dagger(\tau; \tau_0)
  \left[
  i \partial_\tau \delta_{a b}
  -
  h_{a b}(\tau, \tau_0)
  \right]
  A_{b H}(\tau; \tau_0)
  %
  \nonumber \\
   =&
  \sum_{n, m} \int^0_{-\infty}\!\!\!\!\! d \pv\,
  A_{(n, \pv) H}^\dagger(\tau; \tau_0)
  \left[
  i \partial_\tau \delta_{n m}
  -
  h_{n m}(\tau, \tau_0, \pv)
  \right]
  A_{(m, \pv) H}(\tau; \tau_0)
  .
  \label{eq:S_2nd_quant_particle_gen_bck}
\end{align}
where
\begin{align}
  h_{ a b}(\tau, \tau_0)
  =&
  \int d^{D-2} x\,\int^0_{-\infty}\!\!\! d \pv\,
  \tPsi_a^*(\tau, \pv, x)
  H_{lc}(\tau, x, \pv, p)
  \tPsi_b(\tau, \pv, x)
  \nonumber\\
  =\, &
  \delta(k_{v\, a} - k_{v\, b})\,
  \int d^{D-2} x\,
  \tpsi_n^*(\tau, \pv, x)
  H_{lc}(\tau, x, \pv, p)
  \tpsi_m(\tau, \pv, x)
  .
\end{align}
It follows immediately that
\begin{align}
  A_{a H}(\tau; \tau_0)
  =&
  \sum_b \negthickspace\negthickspace\negthickspace\negthickspace\negthickspace\int
  A_{b H}(\tau_0; \tau_0) U_{S, b a}(\tau, \tau_0)
  ,
  %
  %
  %
  %
\end{align}
where $U_{S, b a}(\tau, \tau_0)$ is the Schrödinger evolution
operator in the $\{ \tPsi_a \}$ basis.
Therefore the time evolution does not mix creation and
annihilation operators hence the annihilation operators annihilate the
vacuum for all times.
Said differently, the vacuum would in principle depends on $\tau_0$ and $\pv$ but it is
actually independent because of the time evolution of $A_H$ which does
not involve $A_H^\dagger$.

\subsubsection{Comparing with Light-Cone Field Theory}
If we compare the previous quantum action
 \eqref{eq:S_2nd_quant_particle_gen_bck}
 with the light-cone action of a complex
 scalar field
\eqref{eq:S_2nd_quant_LCFT_gen_bck}
we see they match only when we introduce two particles with opposite charges
so that we can make
the identifications  between the quantum fields
$\thphi_H(u, \kv, x) = \tPsi_{1st\, part\,\, H}(\tau=u, \kv, x)$
and
$\thsphi_H(u, \kv, x) = \tPsi_{2nd\, part\,\, H}(\tau=u, \kv, x)$
for $\kv<0$. 
For the real scalar we need instead one particle 
so that we have 
the match $\thphi_\R(u, \kv, x) = \tPsi_{part\,\,H}(\tau=u, \kv, x)$ 
 for $\kv<0$.

Notice that we have coupled one particle with an
electromagnetic field and therefore we expect the particle to describe
a complex scalar field while it may describe a real scalar which should
not couple to an electromagnetic field. The resolution of this puzzle is that the coupling of a particle with
an electromagnetic background describes something like
$e^- + \gamma^* \rightarrow e^-$ where $\gamma^*$ is a virtual photon.
If we want to describe the process
$e^+ + \gamma^* \rightarrow e^+$ we need another particle with
opposite charge and this means exactly the introduction of another second quantized field; then we can
match the light-cone action for a complex scalar which is the
``double'' of the real one.


\subsection{On the Meaning of the Wave Function $\tpsi(\tau, \kv, x)$}

We will now discuss the meaning of the wave function
$\tpsi(\tau, \kv, x)$ associated with the Hamiltonian
\eqref{eq: H_LC_1st}.
The wording would suggest that this is a usual wave function which can
be interpreted as probability density.
This is not the case since its real meaning is a charge density.
In order to uncover this meaning we discuss and compare the conserved currents
in the particle and scalar field cases.

\subsubsection{Conserved Current for the Particle}

Let us consider the more general Hamiltonian
\eqref{eq:gen_H_1st_quant_particle_gen_bck} and discuss and
construct the conserved ``probability'' current.
If we take two solutions of the Schrödinger equation
\begin{equation}
  i \frac{\partial}{\partial \tau} \tpsi(\tau, \pv, x)
  =
  H_{l c (1st), \mu, \rho} \tpsi(\tau, \pv, x)
  ,
\end{equation}
we can ask whether the quantity 
\begin{align}
  Q(\tpsi_1, \tpsi_2; \tau_0)
  =&
  \int d^{D-2} x\, \mu(\tau_0, x)\, \tpsi_1(\tau_0, x)^* \tpsi_2(\tau_0, x)
\label{eq:particle_conserved_charge}
\end{align}
is conserved.
In particular, when the two $\tpsi$ are the same
$Q(\tpsi, \tpsi)$ can be interpreted as probability being
always non negative.
Using the hemiticity of the Hamiltonian we get
\begin{align}
  Q(\tpsi_1, \tpsi_2; \tau_1)
  -
  Q(\tpsi_1, \tpsi_2; \tau_0)
  =&
  \int^{\tau_1}_{\tau_0} d\tau\,
  \partial_\tau   Q(\tpsi_1, \tpsi_2; \tau)
\nonumber\\
  =&
  \int^{\tau_1}_{\tau_0} d\tau\,
  \int d^{D-2} x\,
  \partial_\tau \mu(\tau, x)\, \tpsi_1(\tau, x)^* \tpsi_2(\tau, x)
  .
\end{align}
Therefore whenever $\partial_\tau \mu(\tau, x)=0$ we get a conserved
charge. 
More explicitly, when $\partial_\tau \mu(\tau, x)=0$
we can introduce the gauge invariant current:
\begin{align}
  \tilde J_i
  =&
  -i \sigma \left(
  \frac{\tpsi_1^*}{\rho} \nabla_i \, \frac{\tpsi_2}{\rho}
  - \nabla_i \frac{\tpsi_1^*}{\rho} \, \frac{\tpsi_2}{\rho}
  \right)
  ,
  \nonumber\\
  \tilde J_v
  =\,&
  2 \pv \tpsi_1^* \tpsi_2
  =   (\pv \tpsi_1)^* \tpsi_2 +   \tpsi_1^* (\pv \tpsi_2)
  .
\label{eq:particle_conserved_currents}
\end{align}
$\tilde J_v$ and $\tilde J_i$ satisfy the continuity equation
\begin{equation}
  -\partial_\tau (\mu \tilde J_v) + \cD^i \tilde J_i
  =0
  ,
\end{equation}
where $\cD_i$ is the covariant derivative \wrt\,  
the ``total'' gauge field $A_i -\sqrt{2} l_i p_v$
and $\bar g_{i j}$, under the assumption that $\tilde J_i$ is a 1-form.
This construction works for any $\mu, \rho$ and $\sigma$, in particular
for the special values \eqref{eq:mu_rho_sigma_scalar} required to
reproduce the light-cone quantization of the scalar field.


\subsubsection{Conserved Current for the Scalar Field and Klein-Gordon Product}
For a complex scalar field we can define as usual the Klein-Gordon current as
\begin{align}
i  J_\mu(\phi_1, \phi_2)
  =\, &
  \phi_1^* \partial_\mu \phi_2 -   \partial_\mu \phi_1^*  \phi_2
  \nonumber\\
  =
  \frac{i}{\sqrt{|\bar g|}}  \hat J_\mu(\hphi_1, \hphi_2)
  =\, &
  \frac{1}{\sqrt{|\bar g|}}
  \left(
  \hphi_1^* \partial_\mu \hphi_2 -   \partial_\mu \hphi_1^*  \hphi_2
  \right)
  ,
\end{align}
which is conserved as
\begin{align}
  D^\mu J_\mu
  &=0
  ,
\end{align}
where $D_\mu$ is the spacetime covariant derivative as given in
\eqref{eq:NO_prel_divergence}. 
For a complex scalar the  meaning of $J_\mu(\phi, \phi)$ is that
the electrical current associated to the obvious $U(1)$ is conserved.
For a real scalar the current $J_\mu(\phi, \phi)$ vanishes identically
but $J_\mu(\phi_1, \phi_2)$ can be used to define the conserved
Klein-Gordon product.

We can now examine the conditions for the existence of equal $u$
conserved charge by computing:
\begin{align}
  &
  \int_{[u_0, u_1]} d u\, \int d v\, d^{D-2} x 
  \sqrt{|\bar g|}\, D^\mu J_\mu
  =
  \nonumber\\
  & =
    \int_{[u_0, u_1]} d u\, \int d v\, d^{D-2} x \,
    \Bigl\{ 
    - \partial_u \left( \hat  J_v \right)
    + \partial_v 
    \left[ - \hat J_u + (\bar l^{\,2} -h) \hat J_v + \bar l^i \hat J_i\right] \nonumber \\
    & \phantom{= \int_{[u_0, u_1]} d u\, \int d v\, d^{D-2} x \,
    \Bigl\{ }
    + \partial_i \left(
      \bar l^i \hat J_v + \bar g^{i j} \hat J_j \right)    
    \Bigr\} =
  \nonumber\\
  & =
  - \int d v\, d^{D-2} x\, \hat J_v |_{u_1}
  + \int d v\, d^{D-2} x\, \hat J_v |_{u_0}
  \nonumber\\
  &\phantom{=\,\,}+
  \int_{u_0}^{u_1} d u\, d^{D-2} x\,
  \left[ - \hat J_u + (\bar l^{\,2} -h) \hat J_v
  + \bar l^i \hat   J_i\right]|^{v=+\infty}_{v=-\infty}
  \nonumber\\
  &\phantom{=\,\,}
  +\sum_{i=2}^{D-2}
  \int_{u_0}^{u_1} d u\, d v\, \frac{d^{D-2} x}{d x^i}\,
  \left(
  \bar l^i \hat J_v + \bar g^{i j} \hat J_j \right)
  |^{x^i=+\infty}_{x^i=-\infty}
  ,
\end{align}
where we have used eq. \eqref{eq:NO_prel_divergence}.
It follows that the charge
\begin{align}
  Q(\hphi_1, \hphi_2)
  =&
  \int d v\, d^{D-2} x\, \hat J_v(\hphi_1, \hphi_2) |_{u_0}
\label{eq:KG_charge}
\end{align}
is conserved if the appropriate boundary conditions are chosen, i.e when
the currents $\hat J$ vanish at ``space'' boundary.
The same condition on the $v$ boundary is necessary to write the
action \eqref{eq:S_2nd_quant_LCFT_gen_bck}
which was obtained by  dropping some boundary terms.

\subsection{The Physical Meaning of the Wave Function $\tpsi(\tau, \kv, x)$}

Given the fact the $\tpsi(\tau, \kv, x)$ follows a Schrödinger
equation and that we can find a non negative conserved density
\eqref{eq:particle_conserved_charge},
it is
natural to think that it can be interpreted as a non relativistic wave
function.
Actually this is not a reasonable interpretation.
The main reason is that the measure used is not the natural and
physical measure.
In fact the natural measure one would like to use from the GR point of
view would be the one derived from the space distance $d l^2$.
This measure is however null since we are on a null surface.
Even forgetting about this point and accepting we can use $d\kv$ as
measure for the partially Fourier transformed wave function
$\tpsi(\tau, \kv, x)$ we have the following problem.
The spacial distance can be defined and measured using light rays
and the volume is $vol= \sqrt{\frac{ |\bar  g|}{ |h|} } d^{D-2} x$.
Taking into account that $\tpsi(\tau, \kv, x)$ is a density one
would like to use $vol= \sqrt{\frac{ 1}{ |h|} } d^{D-2} x$
but this is not the natural measure from the LCQFT point of view.

So how can we interpret $\tpsi(\tau, \kv, x)$?
Looking to the way we have arrived to the 2nd quantized theory, it seems natural
to interpret $\tpsi_a(\tau, \kv, x)$ as a mode for the 2nd quantized
theory which can be read from the one particle amplitude in LCQFT
in the Heisenberg picture
\begin{align}
  \langle \Omega | \tPsi_H(\tau, \pv, x)
  \left(  a^\dagger_{(n,\kv)} | \Omega\rangle \right)
  =
  \tpsi_n(\tau, \kv, x)\, \delta(\pv - \kv)
  .
\end{align}
This approach while technically correct is not very illuminating.
A more physical meaning can be obtained using the conserved current in
LCQFT.
We can use the LCQFT current  since if we compare
the charge density in LCQFT \eqref{eq:KG_charge}
and the 1st quantized charge \eqref{eq:particle_conserved_charge}
we see they essentially match.
Let us evaluate the vev of the LCQFT charge
density in the one particle state.
The normal ordered space and time smeared charge in Heisenberg picture reads:
\begin{align}
  Q_{f H}
  =&
  \int d u\, d v\, d^{D-2} x\,   \sqrt {|\bar g(u, x)|}\,
  f(u, v, x)\,: J_{v H}(u, v, x) :
  \nonumber\\
  =&
  \int d u\, d^{D-2} x\,
  \Biggl[ \, \int_{-\infty}^0 d k_{1 v}\,d k_{2 v}\,
  \frac{\tilde f(u, +k_{1 v} -k_{2  v}, x) }{\sqrt{2\pi}}\,
  \frac{ k_{1 v} +  k_{2  v} }{2 \sqrt{| k_{1 v}  k_{2  v} |} }
  \nonumber\\
  &
  \sum_{n, m}
  a^\dagger_{(n, k_{1 v})}\,   a_{(m, k_{2 v})}\, 
  \tpsi_n^*(u, x; k_{1 v} u_0, e)
  \tpsi_m(u, x; k_{2 v} u_0, e)
  \nonumber\\
  %
  &
  +
  \int_{-\infty}^0 d l_{1 v}\,d l_{2 v}\,
  \frac{ \tilde f(u, -l_{1 v} +l_{2  v}, x) }{\sqrt{2\pi}}\,\,
  \frac{ -l_{1 v} -l_{2  v} }{2 \sqrt{| l_{1 v}  l_{2  v} |} }
  \nonumber\\
  &
  \sum_{n, m}
  :  b_{(n, l_{1 v})}\,   b^\dagger_{(m, l_{2 v})} :\, 
  \tpsi_n(u, x; l_{1 v}, u_0, -e)
  \tpsi_m^*(u, x; l_{2 v}, u_0, -e)
  \nonumber\\
  %
  &
  +
  \int_{-\infty}^0 d k_{1 v}\,d l_{2 v}\,
  \frac{ \tilde f(u, +k_{1 v} -l_{2  v}, x) }{\sqrt{2\pi}}\,
  \frac{ k_{1 v} -l_{2  v} }{2 \sqrt{| k_{1 v}  l_{2  v} |} }
  \nonumber\\
  &
  \sum_{n, m}
  a^\dagger_{(n, k_{1 v})}\,   b^\dagger_{(m, l_{2 v})}\, 
  \tpsi_n^*(u, x; k_{1 v},  u_0, e)
  \tpsi_m^*(u, x; k_{2 v}, u_0, -e)
  \nonumber\\
  &
  +
  \int_{-\infty}^0 d l_{1 v}\,d k_{2 v}\,
  \frac{ \tilde f(u, -l_{1 v} -k_{2  v}, x) }{\sqrt{2\pi}}\,
  \frac{ -l_{1 v} +k_{2  v} }{2 \sqrt{| l_{1 v}  k_{2  v} |} }
  \nonumber\\
  &
  \sum_{n, m}
  b_{(n, l_{1 v})}\,   a_{(m, k_{2 v})}\, 
  \tpsi_n(u, x; k_{1 v}, u_0, -e)
  \tpsi_m(u, x; k_{2 v}, u_0, e)
  \Biggr]
  ,
\end{align}
where $f(u,v,x)$ is the smearing function.
The previous expression implies:
\begin{align}
  \left( \langle \Omega | a_{(m,\pv)} \right)
  Q_{f H}
  \left( a^\dagger_{(n,\kv)} | \Omega\rangle \right) = & \int d u\, d^{D-2} x\, 
  \frac{\tilde f(u, +\pv -\kv, x) }{\sqrt{2\pi}}\,
  \frac{\pv +  \kv }{2 \sqrt{| \pv  \kv |} }
  \nonumber\\
  &
  \tpsi_m^*(u, x; \pv, u_0, e)
  \tpsi_n(u, x; \kv, u_0, e)
  .
\end{align}
When specializing the smearing function to the delta as
$
f_0(u, v, x) = \delta(u -u_0)\,\delta(v -v_0)\,\delta^{D-2}(x -x_0)\,
$
we get the expectation value:
\begin{align}
  \left( \langle \Omega | a_{(n,\kv)} \right)
    \left( \sqrt{|g|} J_{v H} \right)(u_0, v_0, x_0) :
  \left( a^\dagger_{(n,\kv)} | \Omega\rangle \right)
  =&
  - 
  |\tpsi_n(u_0, x_0; \kv,  e)|^2
  ,
\end{align}
which clearly shows that $  |\tpsi_n(u, x; \kv,  e)|^2$ is a charge
density which is independent on the coordinate $v$.
The absence of $v_0$ dependence is due to the choice of taking
$\pv=\kv$ in the bra and ket states, would we have chosen $\pv$ and
$\kv$ different we would have found a $v_0$ dependence.
This is further confirmed by the fact that the antiparticle has
opposite sign charge density.

This prompts the question on how it is then possible that the second quantized particle has a conserved current \eqref{eq:particle_conserved_currents}
even if it is neutral.
The reason is the absence of interactions. In fact without interactions events like $e \gamma \rightarrow  e e e^+$ are not possible and the number of positive and negative charged particles is conserved.

\section{Conclusion}
In this paper we have examined what is the light-cone evolution and quantization on time-dependent backgrounds.
While this has already been used in string theory in an heuristic and effective way, the general picture has not been considered, at least to the best of our knowledge.
The light-cone evolution on time-dependent backgrounds can be characterized in two equivalent ways.
Either as the fact that the light-cone gauge fixed particle has an Hamiltonian which is free of square roots or as the fact that the constant light-cone time hypersurfaces are null.
We have however omitted the discussion of global issues which would lead too far away, exactly as it was done in the discussion of the gravitation in light-cone gauge \cite{Scherk:1974zm}.
The main result from this point of view is that it is possible to perform the light-cone quantization also on backgrounds which do not admit a null Killing vector.

Another point is that the proper fields to be used in light-cone quantization are not the obvious natural fields but some scalar densities for which the ``$p \dot q$" term has a light-cone time-independent coefficient.

We then considered the connection between the second quantized  particle in the light-cone and the associated scalar field theory.
Usually it is taken for granted that one matches the other.
In the light-cone we must supply the further information that the filed associated with the particle is not a scalar but a scalar density.

Finally we showed that the particle wave functions or the corresponding filed modes must be associated to a charge density and not to a particle density.

{\bf Acknowledgments}
We thank I. Sachs and M. Schneider for a discussion on their results.
This research is partially supported by the MUR PRIN contract 2020KR4KN2 “String
Theory as a bridge between Gauge Theories and Quantum Gravity” and by the INFN project ST\&FI “String Theory \& Fundamental Interactions”. 

\printbibliography

\end{document}